# National Ignition Facility (NIF) Control Network Design and Analysis

R. M Bryant, R. W. Carey, R. V. Claybourn, G. Pavel, W. J. Schaefer
LLNL, Livermore, CA 94550, USA

## Abstract

The control network for the National Ignition Facility (NIF) is designed to meet the needs for common object request broker architecture (CORBA) inter-process communication, multicast video transport, device triggering, and general TCP/IP communication within the NIF facility. The network will interconnect approximately 650 systems, including the embedded controllers, front-end processors (FEPs), supervisory systems, and centralized servers involved in operation of the NIF. All systems are networked with Ethernet to serve the majority of communication needs, and asynchronous transfer mode (ATM) is used to transport multicast video and synchronization triggers. CORBA software infra-structure provides location-independent communication services over TCP/IP between the application processes in the 15 supervisory and 300 FEP systems. Video images sampled from 500 video cameras at a 10-Hz frame rate will be multicast using direct ATM Application Programming Interface (API) com-munication from video FEPs to any selected operator console. The Ethernet and ATM control networks are used to broadcast two types of device triggers for last-second functions in a large number of FEPs, thus eliminating the need for a separate infrastructure for these functions. Analysis, design, modeling, and testing of the NIF network has been performed to provide confidence that the network design will meet NIF control requirements.

## 1 INTRODUCTION

The NIF is being developed for laser fusion and high-energy-density experimental studies [1]. NIF will consist of 192 laser beam lines that are focused onto a target within the target chamber. The Integrated Computer Control System (ICCS) is being developed to provide distributed control and monitoring of the approximately 60,000 control points and over 500 video sources in NIF [2]. This integration is provided by approximately 650 computer systems distributed throughout the facility. The ICCS network provides connections for the 16 operator workstations, 300 FEPs, 275 embedded controllers, 40 industrial control and safety systems, 14 distributed workstations, the central process server, the file server, and 150 network outlets distributed throughout the facility for connection of portable computers and diagnostic systems. FEPs and embedded controllers interface to, and are located in close proximity to, the control points. Overall control is orchestrated by the supervisory systems and operator workstations that are centrally located in the NIF computer and control rooms. Real-time control, where required, is provided within the FEPs. A separate timing system provides precise timing signals down to 30-psec accuracy for control and diagnostic systems during the 2-second shot interval [3].

The ICCS is divided into subsystems to partition activity and ensure performance. There are 10 supervisor software applications that conduct NIF shots in collaboration with 17 kinds of FEPs as shown in Figure 1.

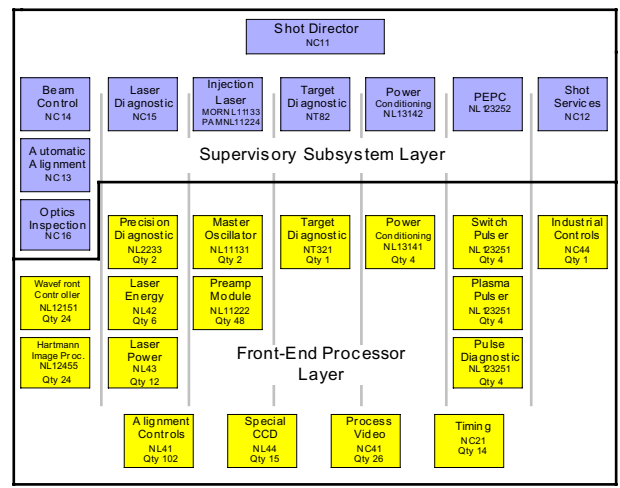

Figure 1 NIF Integrated Computer Control Subsystems

The FEP computers are based on either VxWorks on PowerPC or Solaris on UltraSPARC processors. These systems are primarily diskless and will be bootstrapped over the network.

The supervisory system hardware consists of operator consoles and workstations in the main control room and a central multiprocessor server in the computer room. These systems are based on Sun Solaris platforms. There are eight operator consoles—shot director, industrial controls, laser diagnostics, optical pulse generation, target diagnostics, alignment/ wavefront, power conditioning, and an auxiliary console. Each console consists of two operator workstations with three 20-in. liquid crystal displays each.

The ICCS architecture uses the client-server software model with event-driven communications. It is based on a scalable software framework that is distributed over the supervisory and FEP computers [4]. The framework offers interoperability among different computers and operating systems by using CORBA to provide TCP/IP message transport for interprocess communication.

## 2 NETWORK REQUIREMENTS

Network communication requirements have been identified for each FEP, embedded controller, and supervisory system as a basis for network design. Performance requirements collected included boot image size, initialization message rates and sizes, peak message rates and destination, and shot archive and history data sizes.

The traffic flow is primarily between the centralized Supervisory systems and the distributed FEPs. The majority of network traffic will be asynchronous point-to-point messages that do not have stringent latency or jitter requirements. This includes control messages, sensor data, boot images, archive and history data, etc. This traffic is suitable for standard TCP/IP communication using Ethernet switching technology. The exceptions to this are the requirements for digital video transport and network triggers.

Digital video traffic comes from the 54 video FEPs that interface to the 500 video cameras located throughout the NIF facility. A video FEP can selectively grab single camera images when requested (for example by the automatic alignment system) or can send a stream of frames at up to 10 frames/sec for real-time viewing at the operator workstations. All digitized frames are uncompressed $640 \times 480 \times 8$ bits or roughly 2.5 Mbits per frame. At 10 frames/sec, a video stream requires approximately 25 Mbits/s of bandwidth. Compression is not used because it would impose excessive processing load on the sending and receiving systems, and there is sufficient network bandwidth to support uncompressed video transfer. Each operator workstation is required to display up to two video streams. Each video FEP will be able to source at least two concurrent video streams. The video stream will be multicast when multiple operator workstations want to view the same camera.

Network triggers are short messages that are broadcast or multicast to particular FEPs to trigger specific events just prior to firing the laser. These require a network latency of less than 5 msecs. The FEPs use network triggers to initiate a time-critical function, such as to notify the video FEPs to capture the next video frame and to prepare the alignment control systems for an imminent shot. The video capture trigger is sent over the ATM network across a multicast permanent virtual circuit (PVC). The alignment control system trigger is broadcast over the Ethernet network at 100 Mb/s.

## 3 NETWORK DESIGN AND ANALYSIS

All ICCS systems are networked with Ethernet, which provides for the majority of communication needs (Figure 2). ATM is used for specific video transport. The Ethernet and ATM networks are not directly connected. All systems with ATM connections also have Ethernet connections.

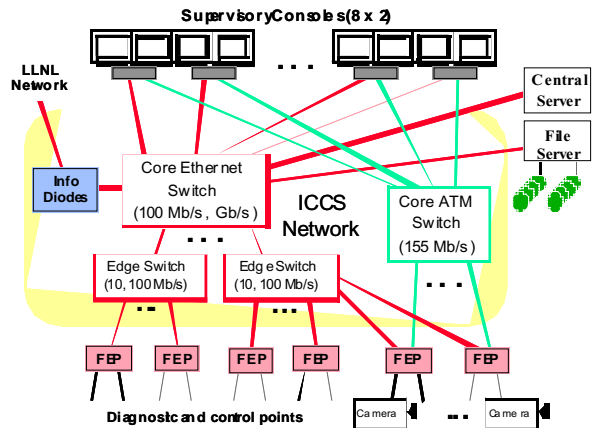

Figure 2: NIF ICCS Network Diagram

The Ethernet network uses switching technology throughout and will be operated as a flat, layer 2 network. Thus, all switching will be based on the Ethernet media access control (MAC) address. This will simplify system configuration and maintainability and will ensure that network triggers on the Ethernet will propagate with a minimum of latency throughout the network. A core Ethernet switch is centrally located in the NIF computer room and provides connectivity to the edge switches that are located thoughout the facility. Fiber cables connect the core switch to the edge switches using 100 Mbit/s Ethernet. With this architecture, there are generally only one or two switches between any two communicating systems, keeping the end-to-end latency low.

The core Ethernet switch is a modular chassis system supporting 100 Mbit/s and Gbit/s. It has layer 2 (MAC) switching and layer 3 (IP) routing capabilities and redundant common logic boards and power supplies for increased reliability. The central server and file server, the most heavily loaded systems, connect directly to this switch using Gbit Ethernet.

Two types of edge switches are used: copper-based 10/100 switches and fiber-based 10 Mbit/s switches. Most of the edge Ethernet switches are 24-port 10/100 units with two 100 Mbit/s multimode fiber uplink ports. The 10 Mbit/s fiber switches connect to systems that require electrical isolation such as the power conditioning embedded controllers. The edge switches connect to the core Ethernet switch over 100 Mb/s fiber links.

Based on the peak Ethernet traffic design requirements for the various FEP and supervisor systems, aggregate message rates and individual system message rates leading up to and following a shot were calculated. Peak aggregate message rate requirements will be less than 2,200 messages per second, which are well within the performance capabilities of current Ethernet switches that provide wire-speed throughput. Individual system message rates will also be well within the capabilities of the network switches and the computer systems.

The ATM network provides OC-3 (155 Mbit/s) connections to the video FEPs and the operator workstations for multicast digital video transport. The ATM network also transports a multicast video trigger from the master timing system to the video FEPs over a multicast PVC.

ATM was selected for its efficiency in transporting digital video. During the network design phase, performance tests were executed to compare Ethernet and ATM performance when sending large video images (307 KBytes/image). A key goal was to minimize the central processing unit (CPU) utilization required for communication on the operator workstations and video FEPs. It was found that using the ATM API directly provided very high throughput with only a small load on the CPUs involved. The key reason for this was the larger message transfer unit (MTU) size available with ATM compared to Ethernet, which reduces the number of packets that need to be processed per image transfer. The multicast video application software uses the XTI API provided with the network interface card.

Simulations were performed using the MIL3 OPNET network modeling tool. Network models were developed consisting of Ethernet and ATM switches with attached supervisory and FEP systems. Customized models were developed for dual-homed systems attached to both Ethernet and ATM. Simulations of the expected latency for the network triggers were performed given various background network loads over Ethernet and ATM. The results verified that the trigger latencies will consistently be well within the 10-msec requirements. This is primarily due to the wire-speed performance of the full-duplex network switches. Simulations of the throughput and latency of a video FEP and operator workstation using the peak network design loading were also performed. The results confirmed that the systems and network will provide the required performance.

## 4 SUMMARY

The ICCS network has been designed using high-performance off-the-shelf switching technology. Testing and analysis indicate that the network design will easily meet the control system throughput and latency requirements for various types of traffic. The network design also provides sufficient capabilities for expansion and higher performance as may be needed in the future.